\documentclass{article}
\usepackage{fancyhdr}
\usepackage{authblk}

\usepackage[english]{babel}

\usepackage[letterpaper,top=2cm,bottom=2cm,left=3cm,right=3cm,marginparwidth=1.75cm]{geometry}

\usepackage{amsmath}
\usepackage{graphicx}
\usepackage[colorlinks=true, allcolors=blue]{hyperref}
\usepackage{listings}

\newtheorem{example}{Example}

\begin{document}

\title{Automatic Generation of Benchmarks and Reliable LLM Judgment for Code Tasks}
%\author{Eitan Farchi, Shmulik Froimovich, Rami Katan, Orna Raz}

\author[1]{Eitan Farchi}
\author[2]{Shmulik Froimovich}
\author[3]{Rami Katan}
\author[4]{Orna Raz}
\affil[1]{farchi@il.ibm.com}
\affil[2]{shmulik.froimovich@ibm.com}
\affil[3]{rami.katan@il.ibm.com}
\affil[4]{ornar@il.ibm.com}
%\date{2024}
\maketitle

\pagestyle{fancy}
%\fancyfoot[CO,RE]{IBM Confidential}
\maketitle

\begin{abstract}
LLMs can be used in a variety of code related tasks such as translating from one programming language to another, implementing natural language requirements and code summarization.  Artifacts generated by state of the art LLM technology are expected to be useful in the sense that a user will be able to use the LLM generated artifact after a small number of easy modifications.  Quantifying this vague notion is challenging and it is thus hard to determine the quality of code related LLM solutions. We refer to evaluation of LLM solutions using LLM judgment as "LLM as a Judge", or LaaJ for short.  In this work we introduce a methodology to generate and evaluate LaaJ implementations, utilizing an automatically generated benchmark.  The purpose of the benchmark is two fold, namely, it is used both to develop and validate the LaaJs and to validate and test the LLM code related solution using the LaaJs. 
To that end, we developed an automated benchmark generation engine, which generates code in multiple programming languages for multiple code related tasks and which serves as the input for LaaJ evaluation.  We utilize a graph representation, G, of the potential code related generations.  The graph vertices are generated artifacts and edges represent possible generations, e.g., the generation of a Java program from its natural language requirements.  Utilizing a chain of LLM agents and G we generate code related artifacts.  Using cycles in G we formulate expectations on the generated artifacts.  
%For example, if a description of the program is used to generate the program and then generate its summary then the summary should be consistent with the original description.  
Taking advantage of these formulated expectations enables the development and testing of reliable LLM judgement for usefulness of the artifacts generated by the solution.  Our approach enables the creation of high quality code task solutions.   
\end{abstract}

\section{Introduction}

The goal of our work is to reliably validate the quality of code related tasks. A solution that embeds the generation of code related tasks is required to produce artifacts, e.g., translate code from one programming language to another, that are useful. Useful here is a vague solution requirement that can be quantified as follows - "the user of the solution will use the generated artifact in her programming, possibly after a small number of minor corrections". While such a requirement may be checked by monitoring a deployed solution, that may be too late in the development cycle of the solution. Instead, it is desired to test the quality of generated artifacts early in the process of the solution development. Furthermore, any changes or updates made to the solution, especially to its LLM components, require reevaluating the quality of the solution. It might be the case that such changes would result in user noticeable degradation of usefulness. 

Reliable early testing of quality requires data that represents the distribution of the users' expected inputs to the deployed solution. For example, a code assistant system may be used for a variety of programming languages and a variety of programming related tasks. 
In order to test the solution early in the development process a sample of representative programs are needed for the different programming languages, e.g., C$\texttt{++} $, Python, Java and COBOL, and code related tasks, e.g., translation, generation, completion, test generation and summarization. While gathering these samples is often straightforward, it is challenging to reliably provide expected results. Usually, human judgement is needed to inspect the solution generated artifacts and determine their usefulness. This approach is labor intensive and proxies to the human judgment are desirable.  

It is thus required to be able to generate data that represents a golden reference, i.e., a useful desired artifact. For example, we would like to generate a COBOL program and its ideal summary from the program specification. In addition, we would like to develop proxies for human judgment with measuring capabilities that can inspect an artifact generated by the solution, e.g., a generated summary of the COBOL program, and determine its usefulness. To that end we use chain of LLMs both for the generation and judgement, which approximates human judgement of whether or not a generated artifact is useful.

The main contributions of our work are:
\begin{itemize}
    \item Defining the functional requirements of the solution. Specifically, defining how to measure usefulness thus defining what usefulness is in a particular solution implementation.
    \item Developing a technique for creating a judge metric as well as labeled data to develop and assess the quality of that judge. The data may be used to assess any potential metric, including existing metrics and non-LLM-based metrics. 
    \item Developing regression benchmarks that are simple in that are easy to understand and debug, yet identify major degradation in usefulness. 
\end{itemize}

%TBC - Orna, seeing your 1.1 I think my related work section can be dropped correct?
\subsection{Evaluation metrics for code LLM tasks}
There exist many metrics for evaluating the quality of code LLM tasks. Ideally, generated code evaluation could rely on \emph{functional correctness}---evaluated by executing code against robust tests \cite{Chen2021Evaluating}---but, this is feasible only for simple tasks. However, executing generated functions is often impractical due to incomplete or unavailable codebases and the difficulty of creating comprehensive test cases. Therefore, the majority of the metrics are {\em reference-based} and require ground truth to which they compare the generated output. Examples of reference-based metrics include BLEU \cite{Papineni2002BleuAM}, ROUGE-L \cite{Lin2004ROUGEAP}, and chrF++ \cite{Popovic2017chrFWH} that treat code as text, focusing on n-gram statistics; CodeBLEU \cite{ren2020codebleu} that incorporates code structure; and CodeBERTScore \cite{zhou-etal-2023-codebertscore} that leverages semantic information in the embedding space. These metrics are faced with a double challenge: it is difficult to provide ground truth at scale, and even if a reliable ground truth solution exists, it is difficult to compare whether a given generation is semantically equivalent to that ground truth. 
There exist {\em reference-less} metrics for evaluating the quality of code LLM tasks. The most notable one is LaaJ. An example is ICE-Score \cite{zhuo2024icescore}. However, it is necessary to assess how well the LaaJ is able to mimic the human judgment, e.g., of usefulness. To do that, it is necessary to have data with human judgment. 

There exist many benchmarks for evaluating the quality of code LLM tasks. Examples include HumanEval\cite{Chen2021Evaluating}, CrossCodeEval\cite{ding2023crosscodeeval}, SWE-bench\cite{jimenez2024swebenchlanguagemodelsresolve}.  However, these benchmarks are static, including their ground truth field. This means that if one wishes to get new data, similar to an existing benchmark, that data needs to undergo a labor intensive process of curation.

Our approach enables generating data at scale, along with labels of expected results. 

\subsection{The essence of our approach}
A key element of our approach is a self consistency capability that provides reliable labels of expected results. Having expected results enables automated large scale validation of the proxies that we develop for usefulness of the generated artifacts.  Specifically, we observe that in the domain of code generation a chain of generated artifacts may "loop on itself", i.e., the last generated artifact may be required to be the same as the first element in the chain.  For example, one may start from a description of a program, generate Java code that implements the program and finally generate a summary that explains what the program is doing. The summary should be specifying the same program that was stated in the initial description.  The coding task domain has an implicit directed graph of possible generations, e.g., $(description, COBOL), (Java, Python)$, etc.  In fact, we are not limited to the above example and can extract different usefulness claims from different loops in the directed graph. For example, We can have the following generation $(Java, summary, COBOL, Java)$  and associate it with the claim that the first and last Java programs should be semantically the same.

\begin{figure}
\centering
\includegraphics[width=\textwidth]{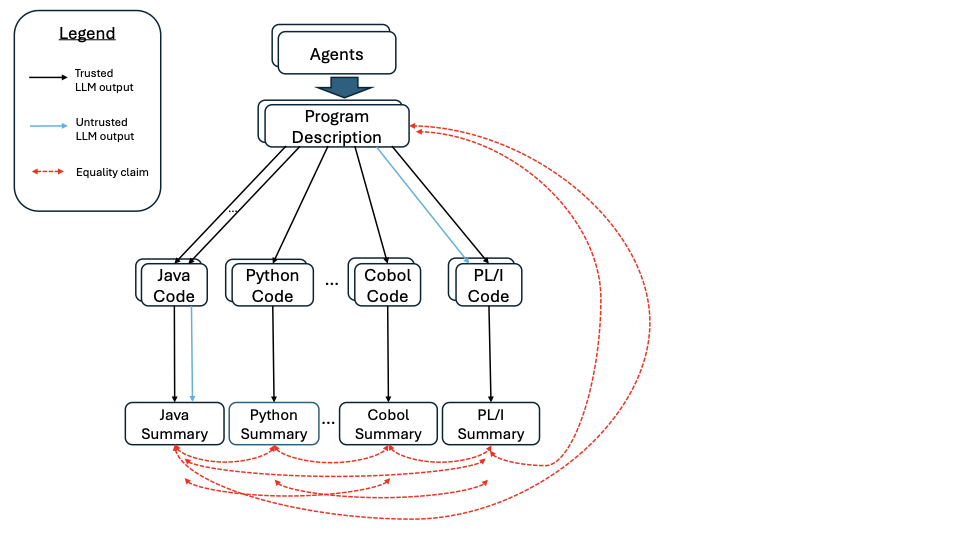}
\caption{\label{fig:selfConsistent} A flow of the self consistency benchmark generation for code generation.}
\end{figure}

Figure \ref{fig:selfConsistent} depicts our approach for creating a benchmark and a proxy for usefulness of code related tasks.  %Next, we explain figure \ref{fig:selfConsistent} top down.  
A chain of agents is used to obtain a set of program descriptions.  The chain includes generating a list of program categories, then a list of program ideas and detailed descriptions for each of the categories.  Each agent in the chain can be enhanced by a human in the loop. 
From the program descriptions an LLM generates: (1) programs for different programming languages that implement each of the descriptions and (2) a natural language summary for each of the programs created in (1).

We introduce the notion of an LLM generation that can be trusted --- depicted by black arrows in Figure \ref{fig:selfConsistent} --- vs. an LLM generation that is being tested and cannot be trusted --- depicted by blue arrows in Figure \ref{fig:selfConsistent}.  Using this notion we can create specific LLM solutions that are geared towards paths in the graph that we are interested in and are more efficient.  
The advantage of the approach is that it provides self checking claims on the solution correctness.  In our running example, given that summaries should be equal we obtain from the axioms of equality, and from careful definition of what equality means in our context, the following ways in which we can self check our LLM judgment. The red arrows in Figure  \ref{fig:selfConsistent} depict claims on that graph.

\begin{enumerate}
    \item Equality is symmetric.  Thus if the LLM judge says that the summary in $(description, Java, summary)$ is equal to the summary in $(description, COBOL, summary)$ then it should also say that the summary in $(description, COBOL, summary)$ is equal to the summary in $(description, Java, summary)$.  Similarity, equality should be transitive obtaining another correctness expectation from the LLM judgment. 
    \item Summaries of the same seed program description should be equal.  For example, the summary that is obtained by $(description, Java, summary)$ and the one that is obtained by $(description, COBOL, summary)$ should be equal.
    \item Summaries obtained by different descriptions should not be equal.  For example, the summary that is obtained by $(description_1, Java, summary_1)$ and the one that is obtained by $(description_2, Java, summary_2)$ should not be equal.
    \item If the two descriptions are of programs that are similar but not the same, e.g., different type of sorting algorithms, the LLM judgment should still be able to distinguish if the summaries are different as long as the summaries are intended to be detailed. Note that this brings forward a key point that has to do with the abstraction level of the summary which we further elaborate on in the following sections.
    \item Summaries should be equal to the original description.  For example, in $(description, Java, summary)$, $description$ and $summary$ should be the same up to implementation details, i.e.,  the summary obtained from the Java program may include details such as variable names that do not appear in the original description of the program.   Our checking should ignore such differences when we determine equality. 
\end{enumerate}

Taking into account the abstractions of the programs we can obtain additional interesting self claims.  For example, summaries of functions should be the same regardless of the context.  This also enable the introduction of stronger tests to the solution by generating larger programs that use some abstractions such as functions.  We refer to this as perturbations on the basic program descriptions and depict it as additional variation in the graph, $G$, of artifacts generation.  

There are several challenges that our approach overcomes. 
As we use the generated benchmark to find problems and expose them to the developers of the solution we should be careful to avoid overfitting of the solution.  To that end we use randomization of generation choices and perturbations of the programs to obtain a fresh instance of the benchmark.  This is done while automatically validating that the LLM judgment is still working. As we develop the LLM judgement we need to avoid overfitting the judgement to a small set of examples.  As the generation process is not limited to a small set of examples and given that we are able to create self checking claims and validate them we are thus able to overcome this pitfall and create self validated LLM judgement.

As testers of the solution we need to be able to determine if a specific execution is useful.  Otherwise, we cannot test the system.   This contradicts the need for realistic long programs that represents real life scenarios.  To address that a balance is obtained.  On that one hand, the descriptions that we obtain are of well defined and simple program requirements which enable the analysis of the solution behaviour.  On the other hand, the generation is built with perturbation capabilities that create more realistic long programs.  In addition, it is still advisable to use another smaller benchmark of real life programs.  Thus, the approach can be viewed as a type of "unit test" of the LLM based solution.

Next we detail our approach for creating self checking benchmarks, using code related tasks as our running example. We note that the approach is general and will be applicable to any context of LLM tasks in which loops in $G$ can be utilized to obtain loop based correctness claims.

The rest of this paper is organized as follows.  We first discuss the requirement definition of an LLM solution for code related tasks (\ref{reference}). Then in \ref{generation} the generation of a benchmark is discussed.  Next LLM judgement is developed in \ref{judge} and analyzed in \ref{judgeAnalysis}.  We discuss pitfalls of the approach and how to overcome them in \ref{pitfalls} and then conclude the paper. 

\section{Quantifying the main requirement - usefulness of code related generated artifacts}
\label{reference}

When developing a solution using an LLM the first thing that needs to happen is defining the requirements of the solution.   These requirements need to be quantified in a way that represents the expected interaction with the solution and the expected solution generated artifacts.  Thus, a definition of desired test data and measures that measure the requirements, i.e., a benchmark for validating the solution, emerges.  

For code related tasks and given the current state of the art, the main requirement is "usefulness".  For example, when generating code from natural language specification, if the code is taken by the developer, modified and then used to develop the final program then the generated code is considered useful.  Also, as mentioned in the introduction we want to focus on testing usefulness early in the development cycle thus we focus on generation of test data and the development of proxy measures to measure usefulness.   

Usefulness here is a tricky requirement which is hard to quantify and its ultimate judge is human judgment.  In addition, the concrete meaning of usefulness changes from one generation task to another.  For example, assume that our code related solution is also required to create a useful summary of a code snippet. Some thought and experimentation leads to the conclusion that a useful summary means a summary that is essentially correct, maps to the original code in a way that helps the user understand the summary and is not unnecessarily lengthy. This leads to the definition of labels such as "correctness", "specific" and "concise" for which a human is the ultimate judge of.  Note the difference in the meaning of usefulness when compared to the code generation task mentioned above.  

Once the requirements are articulated with appropriate labels for different code related generation tasks, we want to obtain labeled data that will let us develop a LLM as a Judge (LaaJ) proxy to usefulness.  In addition, one can also validate and utilize some analytic proxy for usefulness if available.   The LaaJ if successfully validated will serve as a proxy for usefulness and will let us validate the solution at scale which is impossible if we only rely on human judgement.  

If the vague requirement was 'good performance' it would have been relatively easy to quantify it.  In contrast, in our case, usefulness represents human preference and is thus harder to quantify and harder to obtain a proxy for. 

To obtain an LLM judgement for usefulness we first elaborate on the meaning of usefulness in our context and make it concrete.  The description that we create is then used to prompt the LLM and define to the LLM the type of human judgment proxy that it needs to carry.  The developed usefulness description has the additional benefit that it can be used to guide human labeling that in turn will help validate the LaaJ proxy.  

A simplistic approach to judgement would be to create an LaaJ that determines if a generated artifact is useful or not. We hope to get a more refined proxy of LLM judgment that will distinguish between different levels of usefulness of generated artifacts and provide concrete information on why the generated artifact is considered more or less useful.  To that end we further detail usefulness as follows. 

One way that proved helpful in guiding human judgment preference is the introduction of scales. The same scale and its verbal description can be utilized to guide both an LLM and human labelers.  As an example, the below scale may be used for a code summarization task, where a higher score indicates a more useful summary. In our example, the first four categories represent a generated artifact that is not useful while the other three categories represent increasing degrees of usefulness.   %The scale below is given for illustrative purposes.  

\begin{enumerate}
    \item An empty summary.
    \item A completely irrelevant summary or a duplication of the input.
    \item A hallucinated summary that is somewhat related.
    \item The summary is poor and is not useful.
    \item The summary is fair and is at the minimal level of usefulness.
    \item The summary is good but is missing minor elements.
    \item The summary is excellent and adheres to all of the requirements.
\end{enumerate}

%\begin{itemize}
%    \item Code explanation - code snippets are explain in natural language
%    \item Code generation - given a description of what the code is required to do, code is generated to implement the description
%    \item Code completion - the developer writes some code the the solution attempts to complete the code.
%    \item Code translation - the solution attempts to translate the code from one programming language to another
%    \item Test generation - tests are generated to test a given program
%    \item Code remediation - the solution suggests fixes to the given code that removes mistakes or vulnerabilities.
%\end{itemize}

%It is interesting to note that the range of tasks mentioned above are not expected to be perfectly correct. Instead, the solution is expected to provide useful input to the developer.  In each case, the meaning of usefulness changes a bit.  For example, useful code generation or translation means code that can be made correct with a few changes made by the developer, while useful bug fixes will mean that most of the suggested corrections are in principle correct, otherwise a developer will stop using the feature.  Thus, the task of translating and quantifying the usefulness requirement varies cross the AI coding domain.  

%The ultimate judge of whether or not the generated artifact is useful is the developer.  Thus, defining usefulness in principle includes a determination of whether or not the generated artifact possibly after modification by the developer.  

Once a scale is available, there are several ways to obtain labeled data.  One way is to deploy the solution, then monitor and obtain feedback from a human, who in our case is probably a developer, using it.  The clear drawback of doing this is that the testing is done late once the solution is deployed.  Instead, our approach shifts left the validation of the solution by generating the validation benchmark even before the solution exists.  Thus, our approach can be thought of as an instance of the test first approach to software development.  

Another approach to obtaining labeled data is realized by a team of developers that are trained in inspecting and labeling the data for its usefulness.  This approach is labor intensive and does not scale. Labeling at scale is necessary for the validation of LLM based solutions as LLMs are statistical models and their validation requires reasonably large data sets.  
In addition, we found it useful to generate clearly defined and small code snippets.  This lowers the barrier to human labeling thus mitigating the need for a specifically trained team for labeling.   Instead, no special training is needed for inspecting an artifact to determine its usefulness with high accuracy.  Admittedly, for some of the code related tasks samples of realistic large scale labeled business code is required for the data set to be representative and in order to enable final validation of the solution.  We detail how we address this challenge in \ref{pitfalls} below.

\section{Creation of a self consistent benchmark}
\label{generation}
We detail the pipeline used to generate a benchmark for code related tasks and its associated LLM judgment.
%We describe the generation of a benchmark and the development and validation of its associated LLM judgments.  

\subsection{Human provided seed topics for generation}

The initial step includes generating a list of program ideas to be used to generate the benchmarks. To accomplish this, an LLM agent  provide a list of seed concepts to serve as a starting point. A seed concept is then expanded by a LLM into a diverse vector of program ideas that cover it.  As mentioned above, human in loop can be used to enhance each of these steps.  

The seed concept can be formulated in various ways, depending on the desired scope and validation objectives. It may represent a domain of problems, such as computer science logical problems, SQL challenges, or string manipulation tasks. Alternatively, it could pertain to a specific application area, such as financial application ideas or concepts for simple games. The seed may also define a statement-based domain, for example, involving file commands, control flow, or database operations. Ultimately, the seed can encompass any thematic or categorical direction the user aims to explore, providing the basis for subsequent expansion and diversity in the benchmark set.

\subsection{Seed code requirements}

The second step involves expanding the program ideas, using an LLM, into detailed, elaborated descriptions prompts that are sufficiently clear for an LLM to generate the corresponding program. These descriptions must be language-independent to ensure that programs can be generated in any desired programming language. Two examples of such expansions follow. 

\begin{itemize}
\item
Idea: "Anagram Detector - Write a program that checks whether two given strings are anagrams of each other."
\item
Prompt: "Write a program that checks whether two given strings are anagrams of each other. The program should take two strings as input and return true if the strings are anagrams, or false otherwise. Steps: 
\begin{enumerate}
    \item Define a function to check if two strings are anagrams:
\begin{enumerate}
    \item Convert both input strings to lowercase. 
    \item Sort the characters in both strings.
    \item Compare the sorted strings. If the sorted strings are equal, return true, otherwise, return false."
\end{enumerate}
\end{enumerate}

\end{itemize}

\begin{itemize}
\item 
Idea: "Tower of Hanoi Problem - Write a program that solves the Tower of Hanoi problem for n disks."
\item 
Prompt: "Write a program that solves the Tower of Hanoi problem for n disks. The program should take an integer n as input and print the steps to move the disks from the source peg to the destination peg. Steps: 
\begin{enumerate}
    \item Define a recursive function to move n disks from the source peg to the destination peg. 
\begin{enumerate}
    \item Base case: if n is 1, move the disk directly from the source peg to the destination peg. 
    \item Recursive case: move n-1 disks from the source peg to the auxiliary peg, then move the nth disk from the source peg to the destination peg, and finally move the n-1 disks from the auxiliary peg to the destination peg. 
\end{enumerate}
    \item Call the recursive function with the input value of n."
\end{enumerate}
\end{itemize}

This process ensures that each program idea is sufficiently elaborated to allow for seamless generation by the LLM, regardless of the programming language ultimately selected.
%TBC cluster requirememnts to obtain similar programs thus make the jugment stronger
In order to improve the quality of judgments made by the LLM, we define clusters of related program ideas, wherein the programs share certain similarities but are not identical. Having the LLM act as a judge to distinguish between the artifacts (e.g., summaries) of programs within these clusters will lead to more refined judgments, as the comparisons are between related ideas rather than completely unrelated ones. To this end, we ensure that at least 30\% of the seeds are designed to form such clusters, which generate programs within the same conceptual family. For example:

\begin{itemize}
\item
Seed: Sorting Algorithms
\begin{itemize}
\item    
Idea: "Bubble Sort - Write a program that sorts an array using the Bubble Sort algorithm."
\item 
Idea: "Merge Sort - Write a program that sorts an array using the Merge Sort algorithm."
\item 
Idea: "Quick Sort - Write a program that sorts an array using the Quick Sort algorithm."
\item 
Idea: "Insertion Sort - Write a program that sorts an array using the Insertion Sort algorithm."
\end{itemize}
\item
Seed: Graph Traversal Algorithms
\begin{itemize}
    \item 

Idea: "Depth-First Search (DFS) - Write a program that performs Depth-First Search on a graph."
\item 
Idea: "Breadth-First Search (BFS) - Write a program that performs Breadth-First Search on a graph."
\item 
Idea: "Dijkstra's Algorithm - Write a program that finds the shortest path in a graph using Dijkstra's algorithm."
\item 
Idea: "A* Search - Write a program that finds the optimal path in a graph using the A* search algorithm."
\end{itemize}
\item 
Seed: Fitness and Health Applications
\begin{itemize}
    \item 

Idea: "Step Counter - Write a program that tracks the number of steps taken by the user during the day."
\item 
Idea: "Calorie Tracker - Write a program that helps users track their daily calorie intake by logging meals and calculating total calories."
\item 
Idea: "Workout Routine Generator - Write a program that generates a personalized workout routine based on user preferences, such as duration and exercise type."
\item 
Idea: "Water Intake Reminder - Write a program that reminds the user to drink water at regular intervals throughout the day and logs the amount of water consumed."
\end{itemize}
\end{itemize}

This approach ensures that the LLM can provide more nuanced evaluations, leveraging similarities among related programs to produce more accurate and confident judgments. The clustering strategy also helps assess the generalization capabilities of the LLM by evaluating its ability to understand and differentiate nuanced variations of similar programmatic concepts.

\subsection{The code task generation graph G}

%Definition of the code generation graph
We use a code generation multi graph, $G=(V, E, L)$, to generate data, develop LLM judgement and validate LLMs used in the solution. $V$ is a set of code artifacts types and $L=\{LLM_s, LLM_t\}$ are two possible labels of an edge, namely, $LLM_s$ stands for a "strong" LLM that we trust its generation and $LLM_t$ the tested LLM that we are validating which is used in the solution.  The edges are members of $E = V \times V \times L$.  

%Use of the code generaiton graph for generation of data
We use paths labeled by $LLM_s$ to generate data that represents the desired behaviour of the system.  From the previous stage of the generation process we obtain an instance of some $v \in V$, e.g., if $v$ stands for a program description we obtain some specific program description.  We then generate data by following a path that animates from $v$ and has $LLM_s$ as labels, i.e., if $v=v_1, v_2, \ldots, v_n$ is such a path we generate the data by applying $LLM_s$ on $v_1$ to obtain an instance of type $v_2$ and so on. 

%Use of the code generaiton graph of validation of LLM judgmemnt
In what follows $C$ will stand for $COBOL$, $J$ for $Java$, $P$ for $Python$, $D$ for a description that specifies what the program needs to do and $S$ for detailed summary that includes implementation details.
We make correction claims on paths in $G$ the help us develop and validate LLM judgment.   One example of such a claim is associated with loops in $G$.  For example, if we use the loop $(C, J, P, C)$ and generate a specific instance using $LLM_s$, i.e., we start with an instance of a COBOL program snippet, generate using $LLM_s$ a Java instance, then a Python instance and finally a COBOL instance we expect the initial and final COBOL to be the same. This claim can be then used for LLM judgment validation.

Another example occurs when we consider the two paths $(D, C, S)$ and $(D, P, S)$.  In this case, we start from a program description and generate a COBOL program and then its summary or generate a Python program and then its summary. Using $LLM_s$ we expect the two summaries to be consistent up to implementation details.  As before, that claim can be used to validate LLM judgment.  The second example of a correctness claim highlights that correctness here is not necessarily an equivalence claim.  Another example of a correctness claim that is not an equivalence claim will occur when we compose small code snippets to create more realistic large programs (\ref{large}).

%Use of the code generaiton graph for testing of LLM under test
Once LLM judgment is validated $G$ is also used to validate or regress a solution.   In order to do that the following steps are taken.

\begin{enumerate}
    \item A correction claim is picked.  For concreteness assume the correctness claim is the loop 
    $v_1, \ldots, v_n$ $v_i \in V$ in which $v_1 = v_n$.  In addition, assume some of the edges along the loop is of the form $(v_i, v_{i+1}, LLM_t)$.  
    \item We pick an instance of type $v_1$ and generate the loop by applying an LLM on $v_i$ to get $v_{i+1}$.  At least in one of the generation steps $LLM_t$ which is being validated is used. 
    \item Use the LLM judgment developed previously to determine if the instance of type $v_1=v_n$ obtained when we reached $v_n$ in the generation process is the same as the instance of $v_1$ we started with. 
\end{enumerate}

Naturally, if you already generated some of the steps in the the loop above using $LLM_s$ there may be no need to generate them again and only the generation stages that uses $LLM_t$  are applied.  Examples of testing follow. 

\begin{example}
 Assume $G$'s edges are  $(D, C, LLM_s), (C, S, LLM_s),$ 
 $(C, S, LLM_t), (S, D, LLM_s)$ and consider the loop
 $(D, C, S, D)$.  To create a test we pick a description instance and use $LLM_s$ to generate a COBOL snippet and then use $LLM_t$ to create a summary that includes implementation details and finally use $LLM_s$ to generate a description instance which does not include implementation details.  We then use the LLM judgment previously developed to determine if we reached the same description. 
 In this example we are testing if the summary produced by $LLM_t$ is a good summary.   
\end{example}

%TBC provide and example with code snipets

%TBC Shmulik provide an example of correctness claim that is not equality but comparison 
\begin{example}
Assuming there are two initial idea descriptions,, denoted as $D_a$ and $D_b$. These idea descriptions are expanded to generate two corresponding programs, $C_a$ and $C_b$. Subsequently, each program is translated into a different programming language (both $C_a$ and $C_b$ are translated to the same target language, resulting in $T_a$ and $T_b$, respectively). Once translated, both programs are summarized to produce summaries $S_a$ and $S_b$. The final step involves employing an LLM as a judge to compare $S_a$ and $S_b$ with the original description $D_a$, in order to determine which summary more accurately reflects the original idea.
The correctness claim in this scenario is that, when comparing summary $S_a$ and summary $S_b$ to the original description $D_a$, summary $S_a$ will be a better reflection of $D_a$ than $S_b$. 
\end{example}

\subsection{Automatic generation}

In appendix \ref{example} we provide a detailed example of a generation of the following paths $(D, C$\texttt{++} $, S)$ and $(D, Python, S)$ using a strong LLM that can be used as reference.  Indeed, the obtained summaries when striped off the program implementation details are essentially consistent with the initial prompt description specifying the generation.  As a result, the generation below can be used as a reference claim to test weaker but more efficient LLMs that generate, for the current paths, either code from description or summary from code.   

\section{Building and tuning the LLM based judge}
\label{judge}

Constructing an LaaJ is a complex task that primarily involves adjusting the LaaJ for a specific evaluation function. This process of adjustment can be implemented either by tuning the model itself to deliver accurate judgments based on given prompts, or by prompt engineering, crafting specific prompts designed to elicit the desired outcome from the unaltered model.

In many cases, evaluation relies on a predefined scale—such as the 1 to 7 scale for code summarization that Section \ref{reference} described. We define a similar scale below for code explanation. The scale description provides details for assessing the quality of code explanations. In this scale scores 4 and above describe a useful explanation. 

\begin{enumerate}
    \item The explanation is empty or repeats the input.
    \item The explanation is overly abstract, brief and mostly consists of hallucinated content.
    \item The explanation is partial and incorrect, deemed unhelpful and unreliable by experienced programmers for understanding and maintaining the code.
    \item The explanation is incomplete and lacks critical details but provides enough information for an experienced programmer to grasp the general structure of the program.
    \item The explanation includes many useful details but contains some inaccuracies, falling short of enterprise-level standards.
    \item The explanation is missing only minor details, making it understandable and maintainable by a novice programmer.
    \item The explanation is thorough, detailed, and sufficient for a novice programmer to easily understand and maintain the code.
\end{enumerate}
However, this scoring may diverge from human evaluators’ judgments, as the definitions of each level are not deterministic. To externalize the "inner reasoning" of the LLM, we can employ two strategies:

\begin{enumerate}
    \item Ask for reasoning from the LLM used as the judge model so that it would be clearer how the LaaJ evaluates a given explanation. This often results in revealing the linguistic patterns and evaluation criteria that the LLM tends to prioritize.
    \item Present the LLM with an ideal explanation (e.g., one that would score 7 out of 7) and ask for an explanation of why it merits the highest score. Repeating this process for multiple examples (both high- and low-quality summaries) helps refine the set of evaluation categories and the terminology that aligns best with the model’s internal processes.
\end{enumerate}

%TBC: @Shmulik: I think it would be good to have a concrete example for the two items above

These insights can then be applied to craft an optimized scale that is tailored to the specific model, ensuring that the prompt language aligns with the LLM’s inherent evaluation tendencies.

\subsection{Architecture of the LaaJ}

The LaaJ is built utilizing several interconnected layers, which are organized into three main functionalities:

\begin{enumerate}
    \item \textbf{Prompt Optimization}: includes techniques such as dynamic few-shot learning, automated prompt extraction, and model-specific prompt processing, which are designed to enhance the accuracy of the LLM's responses based on the prompt.
    \item \textbf{Optimized Inferencing}: includes managing batch processing, token allocation, and error recovery mechanisms to ensure efficient and reliable performance during inference.
    \item \textbf{Postprocessing}: involves handling the output, including dynamic postprocessing using formatting models, analytical postprocessing, and error handling, to format the results in a deterministic and standardized way that can be seamlessly integrated into downstream systems.
\end{enumerate}
By utilizing these functionality layers, the LaaJ is able to produce consistent, deterministic results, enabling its integration into automated pipelines for scalable and reliable performance.

\section{Measuring and selecting the best judge}
\label{judgeAnalysis}

%TBC - Add coverage of cycles to the white paper.  For any edege generated by $LLM_s$ have a cycle that contains the edge.  Discuss how to obtain a minimal such sets of cycles. 

Once candidate judges, $LaaJ_1, \ldots, LaaJ_k$, are created one needs to choose the best LLM judgment.  In general we may have $l$ correctness claims and we need to choose the best LLM judgment for all of the validation claims.  This is desirable as re-validating the judgment when a fresh sample is generated and introducing minor corrections for a single LLM judgment will be easier than correcting multiple judgments that apply for each correctness claim.  In this work we do not focus on the task of choosing a single LLM judgment, for a large set of validation claims, instead, for simplicity, we assume that only one claim is given, $l = 1$, and focus on the choice of the best LLM judgment in this case. 

To illustrate how choosing the best LLM judgment is done we consider a specific correctness claim.  Given the following three paths in $G$, $(D, C, S)$, $(D, J, S)$ and $(D, P, S)$, the correctness claim we focus on is that the summaries obtained by each of the three paths should be the same up to implementation details.  The generation step previously described using $LLM_s$ gives us a set of data samples, $CLAIM = \{ (S^1_C, S^1_J, S^1_P), \ldots, (S^n_C, S^n_J, S^n_P) \}$ as result of applying the three paths in $G$, $(D, C, S)$, $(D, J, S)$ and $(D, P, S)$ on instances of program deceptions.  $(S^i_C, S^i_J, S^i_P) \in CLAIM$ stands for the $i'th$  summary obtained by the generation using the $i'th$ program description $D_i$ on the three generation paths.  Thus, $S^i_C$ stands for the summary obtained from the generation starting from description $D_i$ on the COBOL generation path $(D, C, S)$.  Indeed, we expect that $(S^i_C, S^i_J, S^i_P)$ are the same summary up to implementation details which instantiates our validation claim and is next used to choose the best LLM judgment.

We thus expect that $LaaJ_l(S^i_{Pr_1}, S^j_{Pr_2}) = True$ if $i = j$ and $False$ otherwise where $Pr_m$ stands for the programming language, $Pr_m \in \{J, C, P\}$.  We give each LLM judgment $LaaJ_l$ a score of $1$ if it judges correctly and $0$ otherwise.  Concretely, the indicator function $I(LaaJ_l, S^i_{Pr_1}, S^j_{Pr_2})$ is $1$ if  $i = j$ and $LaaJ_l(S^i_{Pr_1}, S^j_{Pr_2}) = True$. In addition, $I(LaaJ_l, S^i_{Pr_1}, S^j_{Pr_2})$ is $1$  if  $i \ne j$ and $LaaJ_l(S^i_{Pr_1}, S^j_{Pr_2}) = False$.  In all other cases the indicator function is $0$.  We then consider the average score on $CLAIM$ for $LaaJ_l$ to be $S(LaaJ_l) = \sum_{i, j \in \{1, \ldots, n\}, Pr_m \in \{J, C, P\}, m \in \{1, 2\} } I(LaaJ_l, S^i_{Pr_1}, S^j_{Pr_2}) $.  The LaaJ candidate that obtains the highest score is chosen as the best judge.

%TBC Eitan - add a discussion on the apropriate statisitical test required here. 

\section{A concrete LLM judgment evaluation example}

We attempted to develop an LaaJ for the paths $(D, COBOL, S)$, $(D, PL/1, S)$ and $(D, Java, S)$ with the correctness claim that the summaries should be the same for the same seed if implementation details are ignored but for different seeds the summaries should be different.  

To that end, we conducted an experiment as follows. We began with four manually selected concept seeds, namely, SQL problems, string manipulations, financial application, and computer science logical problems, and used the $LLM_s$ based generation process to create 40 program ideas and corresponding prompts, with 10 ideas for each seed.  We further applied the $LLM_s$ generation process to obtain the code for these prompts, and then a summary for each program language generated code.

Our goal was to perform six comparative evaluations, each involving the comparison of summaries generated for the same program across the three languages, with additional pairwise permutations to check the symmetry of the LLM judgement, resulting in a total of six comparisons. However, due to formatting errors, some programs were corrupted, leaving us with 232 valid pairs of summaries to compare. Each pair consisted of two summaries generated for the same program description but written in different languages. These pairs were assumed to be similar and were therefore labeled as "true."

To counterbalance this, we randomly selected pairs of summaries that originated from different program ideas and labeled them as "false." Each summary pair was evaluated by a Summary-to-Summary Large Language Model as Judge (S2S LaaJ) system, which provided verdicts for all 464 pairs.

The S2S LaaJ was developed in accordance with the methodology outlined in the preceding sections.  We utilized a similarity scale ranging from 1 to 7, where 1 indicated completely different summaries and 7 represented the highest degree of similarity. A threshold score of 4 was set, meaning that scores of 4 or above were deemed similar, while scores below 4 were considered dissimilar. 

Two LaaJ models were evaluated. Both LaaJ models achieved 100$\%$ accuracy in identifying the true and false pairs. Furthermore, when testing the symmetry of the evaluations, i.e., ensuring that $LaaJ(a, b)$ yielded the same result as $LaaJ(b, a)$, we also observed 100$\%$ accuracy.

The distribution of similarity scores for the pairs labeled as "true" across the two LaaJs reveals differences between the two judgment and is outlined in table \ref{tab:similarity-scores}.

\begin{table}[h!]
\centering

\begin{tabular}{|c|c|c|}
\hline
\textbf{Score} & \textbf{LaaJ1} & \textbf{LaaJ2} \\ \hline
1              & 0              & 0              \\ \hline
2              & 0              & 0              \\ \hline
3              & 0              & 0              \\ \hline
4              & 23             & 21             \\ \hline
5              & 44             & 113            \\ \hline
6              & 156            & 85             \\ \hline
7              & 8              & 13             \\ \hline
\end{tabular}
\caption{\label{tab:similarity-scores} Distribution of similarity scores between LaaJ1 and LaaJ2 for pairs labeled as "true".}
\end{table}

\subsection{Experiment results: program idea clusters}

We conducted the same experiment using two clusters of related program ideas, where the programs exhibited certain similarities but were not identical. The two clusters were based on sorting algorithms and graph traversal algorithms. For each cluster, we generated eight program ideas and performed 96 pairwise comparisons, following the methodology outlined previously. The resulting accuracy scores for the two clusters were as follows: laaj1 achieved an accuracy of 98.44\%, while laaj2 achieved 99.48\%.

Below is the distribution of similarity scores for pairs labeled as "true":

\begin{table}[h!]
\centering
\begin{tabular}{|c|c|c|}
\hline
\textbf{Score} & \textbf{laaj1} & \textbf{laaj2} \\ \hline
1 & 0 & 0 \\ \hline
2 & 0 & 0 \\ \hline
3 & 0 & 0 \\ \hline
\textcolor{black}{4} & \textcolor{red}{3} & \textcolor{red}{1} \\ \hline
5 & 2 & 59 \\ \hline
6 & 89 & 34 \\ \hline
7 & 5 & 3 \\ \hline
\end{tabular}
\caption{Distribution of similarity scores for pairs labeled as "true". The scores highlighted in red represent those that were incorrectly labeled as "true."}
\end{table}

When combining the entire benchmark, which involved comparing 29\% of the summaries to other summaries within the same cluster, we achieved final accuracy scores of 99.54\% for laaj1 and 99.85\% for laaj2.

We tried the experiment with different LLMs and found that the consistency of summary similarity was retained.

\section{Pitfalls and how we address them}
\label{pitfalls}
%TBC Eitan - continue reading from here

The approach to the validation of code related solutions presented here has several pitfalls which we next highlight and address.  

\subsection{Overfitting}

%Once problems are identified using generated artifacts that were labeled we can not use the same data points for regression of the solution as these examples may have been used to tune the LLM. 
LLMs are statistical models and as such once we find a problem in the LLM based solution and communicate it to the team working on the development of the LLM solution, we run the risk of an information leak.  As the information we provide may be used in the tuning of LLM models we may end up overfitting the LLM to the examples we provided.  The end result is that we may get a false sense of progress if we keep regressing and validating the LLM solution using the same set of data that revealed the issues in the solution and was exposed to the development team.  

To mitigate this concern we can limit the communication of the problem to the development team by only communicating a generic statement of the problem and by avoiding the sharing of the specific data that revealed the problem.  In addition, periodically, we can regenerate a fresh sample of the data set and re-validate the LLM judgment that we developed on the new generated data set.  The second option is gated by the cost, mainly of LLM inference time, of generating a fresh sample of the data. 

\subsection{LaaJ is only a proxy}

%Even with well understood artifacts, inspecting them is labor intensive and we need to automate the process.  Our approach is to create LLM as a judge (LaaJ) proxies to that end.  We need to reliably validate that the LaaJ accurately capture usefulness of the generated artifacts. 

LLM judgement is only a proxy to human judgment.  Thus, direct eyeballing of the data is continually needed to validate and refine our analysis of the LLM based solution performance.  When eyeballing the data, one should sample a small set of data from the generated data set to avoid biased conclusions about usefulness.  As we are estimating the usefulness of a generated artifact and based on the central limit theorem, 20-30 points should be randomly sampled from the dataset.  Our approach democratize the eyeballing or labeling process by creating a dataset that is well defined and simple so that any one can quickly review a small sample of data points and determine the usefulness of $LLM_t$ generations.  Our experience shows that such reviews can be carried out within half a day to a day of work for tasks such as code summarization or code translation of simple well defined program descriptions.  This helps further validate our analysis of the LLM solution in a way that is efficient and statistically sound.

\subsection{Creation of more realistic but easily readable larger programs}
\label{large}

%TBC Demonstrating that well understood simple programs are useful does not necessarily mean that full scale program generation will be useful.
%TBC Testing the LLM based solution for only small and simple well understood programs does that mean the the LLM based solution generated artifiact 
%Programs should be large realistic business solution for some of the tasks

 Demonstrating that well understood simple programs are useful does not necessarily mean that the LLM solution will work on large and realistic programs.  To mitigate this concern we compose large more realistic programs from smaller programs.   Specifically, we create a table that points to the small well understood programs obtained in the first generation process described above and call them in some order.  

%TBC Correctness claims for the development of LLM judgment is made here on similarity between the summary of components used to compose the larger programs
 
 The correctness claims should be adjusted to take into account the composition step and new LLM judgment should be developed and validated.  An example of such an adjustment follows.  Consider the following two generations $(D, C)$ and $(D, J)$. Assume that we have $k$ descriptions 
 $D_1, \ldots, D_k$ we first generate $k$ well defined and simple COBOL functions $C_1, \ldots, C_k$ and $k$ well defined and simple Java programs $J_1, \ldots, J_k$.  This is done using the trusted LLM, $LLM_s$.  We then compose the $2k$ programs to obtain a large COBOL program $C(C_1, \ldots, C_k)$ and a large Java program $J(J_1, \ldots, J_k)$.  We obtain a summary of  $C(C_1, \ldots, C_k))$, $S(C)$, and a summary of $J(J_1, \ldots, J_k)$, $S(J)$, using $LLM_t$.  We expect that the parts of the summary, $S(C)$, that discusses the functionality of each of the function $C_i$ will be the same as the part of the summary, $S(J)$, that discusses the function $J_i$ up to implementation details.  Using that expectation we can develop LLM judgment for the large composed programs. 

\subsection{Coverage}

%The data should represent the range of usage we expect the solution to be used in.

We would like the data being generated to be representative.   In other words, it should contain samples of the different type of programs, programs specifications, etc, that are used in practice by developers.  One way to approximate meeting this coverage requirement is to ensure that the list of categories we obtain from the LLM in the first stage of the data generation is refined, developed and validated to make it as complete as possible.  This is done by adding categories  based on human knowledge and prompting the LLM to complete the list of categories by introducing examples that are not in any of the categories the LLM suggested. 

\section{Conclusion}

We presented a technology specifically designed to enable the construction and validation of LaaJs through automated data generation. The technology serves as a foundational tool for systematically building and refining LaaJs, offering a controlled environment for evaluating the performance of judgment models on code generation tasks.  Using a graph representation, $G$, to map the spectrum of potential code outputs, we employ a series of LLM agents to create code-related artifacts.  By incorporating cycles within $G$, we generate expectations for these artifacts which act as benchmarks for assessing and improving the reliability of LaaJ models. This approach allows us not only to generate high-quality task solutions but also to iteratively optimize LaaJs, ensuring they effectively judge the utility and relevance of generated artifacts. Through this technology, a robust framework for building LaaJs with consistent accuracy and adaptability across various code generation tasks is created.

Future work will include the development of refined LLM judgment by the generation of graph $G$ perturbations that capture the desired refined validation requirements.

%We presented a technology for generating, tuning, and evaluating LaaJ implementations, and for selecting the best LaaJ for 

%automated data generation used both for evaluating metrics, including

%and for evaluating the underlying LLM-based solution in the context of code generation tasks. We utilize a graph representation, $G$, of the potential code related generations. Utilizing a chain of
%LLM agents and $G$ we generate code related artifacts. Using cycles in $G$ we formulate expectations
%on the generated artifacts. Taking advantage of this formulated expectations enables the development and testing of reliable LLM judgement for usefulness of the artifacts generated by the solution. Our approach enables the creation of high quality code task solutions.

%\include{relatedWork}

\bibliographystyle{alpha}
\bibliography{main}

\appendix

\section{Automatic generation}
\label{example}

Here are two detailed examples for path generation that use the same idea as a seed.

\begin{verbatim}

1.	Seed computer science logical problems
2.	Idea: Matrix Multiplication - Write a program that multiplies two matrices
3.	Prompt: Write a program that multiplies two matrices. 
The program should take two matrices as input and return the
product matrix. Steps: Define a function to multiply two matrices. 
Check if the input matrices can be multiplied (i.e., 
the number of columns in the first matrix matches the number of rows
in the second matrix). Initialize the product 
matrix with zeros. Loop through each element in the product matrix. 
For each element, calculate the dot product of the 
corresponding row in the first matrix and the corresponding column 
in the second matrix. Store the result in the 
product matrix. Return the product matrix.
4.	C++ code: 
#include <iostream>
#include <vector>

std::vector<std::vector<int>> 
multiplyMatrices(const std::vector<std::vector<int>>& matrix1, 
const std::vector<std::vector<int>>& matrix2) {
    int rows1 = matrix1.size();
    int cols1 = matrix1[0].size();
    int rows2 = matrix2.size();
    int cols2 = matrix2[0].size();

    if (cols1 != rows2) {
        throw std::invalid_argument("Matrices cannot be multiplied");
    }

    std::vector<std::vector<int>> productMatrix(rows1, std::vector<int>(cols2, 0));

    for (int i = 0; i < rows1; ++i) {
        for (int j = 0; j < cols2; ++j) {
            for (int k = 0; k < cols1; ++k) {
                productMatrix[i][j] += matrix1[i][k] * matrix2[k][j];
            }
        }
    }

    return productMatrix;
}

int main() {
    std::vector<std::vector<int>> matrix1 = {{1, 2, 3}, {4, 5, 6}};
    std::vector<std::vector<int>> matrix2 = {{7, 8}, {9, 10}, {11, 12}};

    std::vector<std::vector<int>> product = multiplyMatrices(matrix1, matrix2);

    for (const auto& row : product) {
        for (int val : row) {
            std::cout << val << " ";
        }
        std::cout << std::endl;
    }

    return 0;
} 
Summary for C++ code:
Here is a detailed summary of the provided C++ program:

**1. Business Purpose:**
The program is designed to perform matrix multiplication, a fundamental 
operation in linear algebra. It takes two matrices as input and produces 
their product as output. This operation is commonly used in various fields 
such as physics, engineering, computer graphics, and machine learning.

**2. Inputs and Outputs of the Program:**

* Inputs:
	+ Two matrices, `matrix1` and `matrix2`, represented as 2D vectors of integers.
* Outputs:
	+ The product matrix, represented as a 2D vector of integers, 
 resulting from the multiplication of `matrix1` and `matrix2`.

**3. Detailed Functional Summary per Function:**

### `multiplyMatrices` Function:

* Purpose: This function performs the matrix multiplication operation.
* Parameters:
	+ `matrix1` and `matrix2`: The two input matrices to be multiplied.
* Return Value: The product matrix resulting from the multiplication of `matrix1` and `matrix2`.
* Functionality:
	1. It first checks if the number of columns in `matrix1` is equal 
 to the number of rows in `matrix2`. If not, it throws an `invalid_argument` 
 exception, as matrix multiplication is not possible in this case.
	2. It creates a new 2D vector, `productMatrix`, with the correct 
 dimensions to store the result of the multiplication.
	3. It performs the matrix multiplication using three nested loops:
		* The outer loop iterates over the rows of `matrix1`.
		* The middle loop iterates over the columns of `matrix2`.
		* The inner loop performs the dot product of the corresponding 
  row of `matrix1` and column of `matrix2, accumulating the result in the 
  corresponding element of `productMatrix`.
	4. Finally, it returns the `productMatrix`.

### `main` Function:

* Purpose: This is the entry point of the program, demonstrating the usage 
of the `multiplyMatrices` function.
* Functionality:
	1. It defines two example matrices, `matrix1` and `matrix2`.
	2. It calls the `multiplyMatrices` function, passing `matrix1` 
 and `matrix2` as arguments, and stores the result in the `product` variable.
	3. It prints the elements of the `product` matrix to the console, 
 using nested loops to iterate over its rows and columns.

Overall, the program provides a simple and efficient implementation of 
matrix multiplication, with a clear and concise structure.

Python code:
def multiply_matrices(A, B):
    if len(A[0]) != len(B):
        raise ValueError("Matrices cannot be multiplied")

    product = [[0 for _ in range(len(B[0]))] for _ in range(len(A))]

    for i in range(len(A)):
        for j in range(len(B[0])):
            for k in range(len(B)):
                product[i][j] += A[i][k] * B[k][j]

    return product

# Example usage:
A = [[1, 2, 3], [4, 5, 6]]
B = [[7, 8], [9, 10], [11, 12]]
print(multiply_matrices(A, B))
Summary for Python code:
Here is a detailed summary of the provided Python program:

**1. Business Purpose:**
The business purpose of this program is to perform matrix multiplication, 
a fundamental operation in linear algebra. Matrix multiplication is used 
in various fields such as physics, engineering, computer graphics, and 
machine learning. This program can be used in applications where matrix 
multiplication is required, such as:

* Linear transformations
* Data analysis
* Image processing
* Neural networks

**2. Inputs and Outputs of the Program:**

* **Inputs:**
	+ Two matrices, A and B, represented as lists of lists in Python.
* **Outputs:**
	+ The product of the two input matrices, also represented as a list of lists.

**3. Detailed Functional Summary per Function:**

* **`multiply_matrices(A, B)` function:**
	+ **Purpose:** This function takes two matrices, A and B, as input 
 and returns their product.
	+ **Input Validation:** The function checks if the number of columns in 
 matrix A is equal to the number of rows in matrix B. If not, it raises a ValueError 
 with the message "Matrices cannot be multiplied".
	+ **Initialization:** The function initializes an empty matrix, `product`, 
 with the same number of rows as matrix A and the same number of columns as matrix B. 
 The elements of the `product` matrix are initialized to zero.
	+ **Matrix Multiplication:** The function performs the matrix multiplication 
 using three nested loops:
		- The outer loop iterates over the rows of matrix A.
		- The middle loop iterates over the columns of matrix B.
		- The inner loop iterates over the elements of the current row of matrix 
  A and the current column of matrix B, performing the dot product.
	+ **Return Value:** The function returns the `product` matrix.

In the example usage, the program multiplies two matrices, A and B, and prints the result. 
The matrices A and B are defined as:

A = [[1, 2, 3], [4, 5, 6]]
B = [[7, 8], [9, 10], [11, 12]]

The output of the program will be the product of these two matrices.
\end{verbatim}

\end{document}